\documentclass[aps,pre, twocolumn, amsmath,amssymb ,superscriptaddress]{revtex4}

\usepackage[latin1, ansinew]{inputenc}
\usepackage{graphicx}
\usepackage{dcolumn}
\usepackage{bm}

\begin{document}

\title{Root growth and force chains in a granular soil}

\author{Mahmoud Fakih\footnote{presently at Department of Civil Engineering, Lebanese 
American University, Beirut, Lebanon.}}\email{mahmoud.fakih@lau.edu.lb}
\affiliation{LMGC, Universit\'e de Montpellier, CNRS, Montpellier, France.}
\affiliation{AMAP, CIRAD, CNRS, INRA, IRD, University of Montpellier, France.}
\author{Jean-Yves Delenne}\email{jean-yves.delenne@umontpellier.fr}
\affiliation{IATE, INRA, CIRAD, SupAgro, University of Montpellier, France}
\author{Farhang Radjai}\email{franck.radjai@umontpellier.fr}
\affiliation{LMGC, Universit\'e de Montpellier, CNRS, Montpellier, France.}
\affiliation{$\langle MSE \rangle^2$, UMI 3466 CNRS-MIT, MIT Energy Initiative, USA}
\author{Thierry Fourcaud}\email{thierry.fourcaud@cirad.fr}
\affiliation{AMAP, CIRAD, CNRS, INRA, IRD, University of Montpellier, France.}

\date{\today}

\begin{abstract}
Roots provide basic functions to plants such as water/nutrient uptake and anchoring in soil. 
The growth and development of root systems contribute to colonizing the surrounding soil 
and optimizing the access to resources. 
It is usually admitted that the variability of plant root architecture results from the 
combination of genetic, physiological and environmental factors, in particular soil 
mechanical impedance. However this last factor has never been investigated at the soil grain scale. 
In this paper, we are interested in the effect of the disordered texture of granular soils 
on the evolution of external forces experienced by root cap. We introduce a numerical model 
in which the root is modeled as a flexible self-elongating tube that probes a  soil composed 
of solid particles. By means of extensive simulations, we show that the forces exerted on the 
root cap during its growth reflect interparticle force chains. 
Our extensive simulations also show that the mean force declines exponentially with root flexibility, 
the highest force corresponding to the soil hardness. Furthermore, we find that this functional dependence 
is characterized by a single parameter that combines the granular structure and 
root bending stiffness. This finding will be useful to further address the biological issues of 
mechanosensing and thigmomorphogenesis in plant roots.

\end{abstract}

\pacs{45.70.-n,83.80.Fg,61.43.-j} 
\maketitle

\section{Introduction}
\label{sec:intro}

The plant-root system displays two main functions essential for plant 
growth, which  are mechanical anchoring and water and nutrient 
uptake \cite{Lynch:2007aa,HodgeBertaDoussanEtAl2009}. Root growth and development, including branching, 
allow the plant exploring and colonizing the surrounding soil and optimizing the access 
to resources. These processes are driven by genetic, biotic and abiotic factors. 
The observed variability of root architecture, defined as the time-space evolution 
of the root system geometry and topology, can consequently result from the combination of 
endogenous instabilities \cite{Forde2009a} and environmental heterogeneities. Gravitropism, i.e. the growth 
response of plants to gravity, is also an important aspect of root's design \cite{Dupuy2010}. 
The local effect of water and nutrient content on root growth and branching has 
been investigated by several authors \cite{bengough12,valentine2012,HoMcCannonLynch2004,DrayeKimLobetEtAl2010}. 
Some of them used numerical process-based models to analyze and quantify 
water flows within a root-soil domain at the root and plant scales 
\cite{DoussanPierretGarriguesEtAl2006,DoussanPagesPierret2003,ClausnitzerHopmans1994}. 

It is well known that the soil mechanical impedance 
is also an important factor that can affect root growth 
and hence crop productivity \cite{Bengough06}. 
Experimental studies have been carried out in order to characterize and quantify 
this mechanical impact in different soil types at the level of a single root \cite{Bengough99, Kolb2012, Eavis1969} 
or at the plant scale \cite{DeSmet,Becel2012}, sometimes considering the combined effects 
of soil strength and water content on the development and anchorage of roots
\cite{Bengough2011,Dexter1987}. Some of these studies \cite{BengoughKerby2002,YangDefossezDanjonEtAl2017} 
were supported by numerical simulations using the finite element method where the soil was simplified as a 
continuous domain \cite{UpadhyayaRosaWulfsohn2002}. Such simulation approaches are, however, not able to capture grain  
scale effects that can have significant impacts on root growth trajectory and root physiological 
responses. The integration of such small scale phenomena at the level of the whole 
root system can partly explain their architectural variability. 

Representing the complex interactions between a single growing root and soil 
deformation requires a grain-scale approach where
the inhomogeneous texture of the soil can be explicitly described. 
The packing fraction (volume fraction of particles) and interstitial pore distribution 
may influence the root growth via their effects on force transmission, soil strength, aeration 
and hydraulic permeability \cite{valentine2012,bengough12,Materechera92,MacKenzie09}.  
The packing texture determines the maximum pressure exerted 
on the root \cite{Gregory2006}. This value should be compared with threshold 
pressures associated to particular physiological responses, e.g. trajectory deviation 
due to an active movement of the root (tropism) or full arrest of the root
elongation \cite{WhalleyDexter93,Tracy2011,Gregory2006,Clark2003}. 

In the same way, the frictional resistance to sliding  
increases the overall resistance of the soil to penetration, making of the root-soil 
friction angle an important parameter \cite{BengoughKerby2002}. 
Conversely, depending on the root flexibility and diameter, the growth of a  
root inside a ganular soil tends to modify the soil texture.  
Bengough and al. \cite{Vollsnes2010} studied the effect of a growing root 
on the particle displacements within a sandbox by means of the PIV technique. 
They found that the displacements decrease with distance from the 
root over a long distance. Hence, the presence of growing roots or dead roots inside a soil 
affects the growth conditions of each root via their footprint on the packing fraction and 
pore distributions. 

The force that a root must exert to penetrate the soil can be measured at the radicle initiation by measuring 
the force exerted on the seed \cite{Barley,BENGOUGH1990}. 
Once the root has penetrated a distance about six times its diameter into soil, 
it gets anchored and therefore the force transmitted back from the tip along 
the root to the seed is no more equal to the force required to penetrate the 
soil \cite{Bengough1994,Pfeffer93}.  
Attempts of measuring the force acting at 
root tip have been made in laboratory in order to evaluate the maximum growth pressure for different 
root species  \cite{Kuzeja2001,Clark2003,MisraDexter}. The 
radial growth pressure has also been investigated by means of 
photoelastic grains \cite{Kolb2012}. However, such measurements are difficult or 
even impossible to perform in larger scale experiments or natural conditions.

At the grain scale, the root cap interacts with the highly inhomogeneous texture of 
the soil. A flexible and narrow root may easily explore the pore space without being able 
to dislodge the grains equilibrated by the normal and frictional forces exerted by their 
neighboring grains. But there is always a large number of grains that are subjected to weak 
forces and belong to what has come to be known as the 
`weak force network' \cite{Radjai1998,Radjai2015}. As the root rigidity increases, 
the force necessary to bend a root during its growth may overcome large  
interparticle forces and the strong force chains observed in granular 
materials \cite{Liu1995a,Radjai1996,Voivret2009}. 
Hence, for understanding the effect of soil inhomogeneity on the root growth and 
architecture, a key issue is to determine how the interparticle forces and force chains are reflected 
in the forces experienced by the root cap.      

In this paper, we investigate the distribution and evolution of forces exerted by soil grains on    
a root modeled as a self-elongating elastic tube of 
constant thickness. Active apex movements such as gravitropism or other deviations 
due to anisotropic cell growth are not considered here, i.e. the root path is only driven 
by external forces exerted by the surrounding grains. This choice was made in order to 
dissociate the external mechanical component  from the biological component of 
root trajectory. The modeling framework is the Discrete Element Method (DEM) 
in which the soil is represented by a granular material composed of rigid 
grains \cite{CundallStrack1979,RadjaiAndDubois2011,RadjaiJean-Yves2011,delenne2011}. 
The root is characterized by its longitudinal, tangential and bending stiffnesses, as well as its diameter 
and growth rate. We consider different values of the root bending stiffness 
and granular samples of different values of packing fraction and particle size distribution. 
As we shall see, the root-particle 
interaction force has a well-defined probability density, and the mean force 
normalized by the hardness of the granular material 
is governed by a robust scaling as a function of the root bending stiffness.          

In the following, we first describe the root model and numerical procedures 
in Section \ref{sec:num}. In Section \ref{sec:force}, we analyze the probability density function 
of the forces exerted on the root cap. Section \ref{sec:bending} presents the 
evolution of the mean force with root flexibility. Then, in Section \ref{sec:soil}, we 
consider the effects of soil properties such as packing fraction and cohesive stresses on 
the mean force. We conclude in 
section \ref{sec:conclu} with a summary of the main findings of this work and a 
brief discussion of its possible extensions.

\section{Numerical model and procedures}
\label{sec:num}


The general modeling framework for particle dynamics applied to 
a collection of solid particles is the Discrete Element Method (DEM) 
\cite{CundallStrack1979,RadjaiAndDubois2011}. In this method, the particle motions 
are computed by stepwise integration of the equations of motion 
for all particles by accounting for their frictional contact interactions and 
the gravity and boundary forces. We use the same framework to model 
the root, defined as a flexible tube represented by an array of 
segments interconnected by linear springs and growing  
as a result of its continuous elongation. The general numerical model and simulations 
are in two dimensions as the root growth requires long-time simulations and 
we need to perform a large number of independent simulations in order to 
be able to assess the variability of the results and the effect of 
system parameters. We present below in more detail 
the particle interactions and our root model.       

The soil particles are 2D disks interacting via  a 
reaction force $\vec f$ defined by its two components $f_n$ and $f_t$ along and perpendicular to 
the contact normal $\vec n$, respectively: 
\begin{equation}
f=f_n \thinspace\vec n+ f_t \thinspace\vec t
\end{equation}
where $\vec t$ is the unit vector perpendicular to $\vec n$. 
The normal force is the sum of three different forces: 
\begin{equation}
f_n=f_n^e+f_n^d+f_n^c, 
\label{eq:Grains_interaction}
\end{equation}
where $f_n^e$ is the repulsive elastic force, $f_n^d$ is a viscous damping force and 
$f_n^c$ is a cohesion force induced by solid or liquid bonding between the particles. 
The elastic normal force is assumed to be a function of the contact deflection $\delta_n$, 
assumed to be equal to the overlap between the disks for small contact deformations. 
We use a linear elastic law as far as there is an overlap between the particles whereas the force 
vanishes when there is no contact:   
\begin{equation}
f_n^e=
\begin{cases}
      -K_{n}\delta_n  \qquad   \mbox{ if}  \thinspace\thinspace \delta_n < 0  \\
      0 \qquad \qquad  \mbox{otherwise}, 
\end{cases}
\end{equation}
where $K_n$ is the contact stiffness. 

In the DEM particle dynamics simulations, it is necessary to introduce also viscous 
damping to account for energy dissipation due to inelastic collisions between particles.  
To make the energy restitution coefficient independent of the particle mass, the damping force is assumed 
to depend also on the spring stiffness and particle mass as follows:
\begin{equation}
f_n^d=
\begin{cases}
      -2\alpha_n\sqrt{mK_n}{\dot\delta_{n}}  \qquad  \mbox{if}   \, \, \dot\delta_n < 0  \\
      0 \qquad \qquad \qquad \, \, \, \, \, \,\, \, \, \, \, \mbox{otherwise}, 
\end{cases}
\end{equation}
where $m = m_im_j/(m_i+m_j)$ is the reduced mass of the two touching 
particles i and j, and $\alpha_n$ is the damping constant varying between $0$ and $1$. 
The cohesion force $f_n^c$ can have different expressions depending on the source of 
cohesion \cite{Richefeu2006,delenne2002}. 

For the friction force $f_t$, we use an elastic-regularized Coulomb law \cite{Shafer1996,Dippel1997}:
\begin{equation}
f_{t}=
\begin{cases}
      -sgn(\dot{\delta_t}) \mu f_{n}  \qquad  \mbox{if} \, \, \vert \hat{f_t} \vert \geq \mu f_{n} \\ 
      K_{t}\delta_{t}+\gamma_{t}\dot{\delta_{t}}  \qquad \, \,  \,  \,   \mbox{otherwise}, 
\end{cases}
\end{equation}
where
$K_t$ is the stiffness of the tangential spring, $\gamma_{t}$ is the tangential damping constant, 
$\delta_{t}$ is tangential spring elongation since the contact between two particles is established, 
$\dot\delta_{t}$ is relative tangential velocity and $\mu$ is the friction coefficient.

\begin{figure}
\centering
\includegraphics[width=0.7\columnwidth]{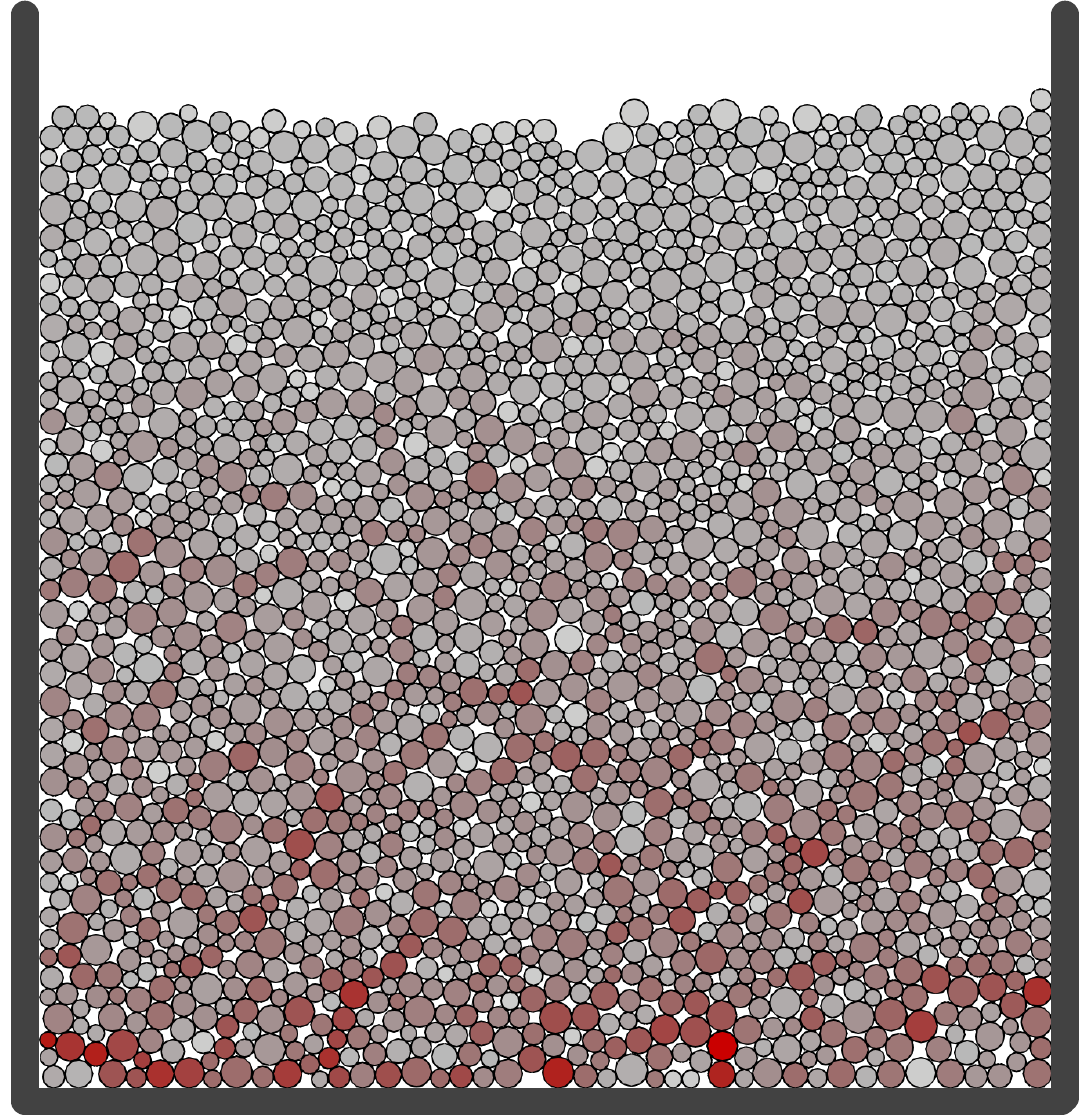}
\caption{(color online) A granular sample at the end of relaxation to equilibrium. Color intensity is proportional to 
average normal force supported by each particle.}
\label{fig:grains}
\end{figure}

The granular samples are prepared by sedimentation: The particles are placed  
on the nodes of a triangular lattice without overlap and allowed to fall into a 
box under the action their own weights. The final state after sufficient relaxation of the particles 
is a dense packing in which all forces on all particles are balanced.   
A weak size polydispersity of particles is introduced to avoid long-range ordering of the disks in 2D. 
The particle radii $R_p$ are distributed between $R_{min}$ = 1 mm, and $R_{max}$ = 2 mm 
with a uniform distribution of their volumes. This distribution ensures that the 
total volumes of the particles in all particle size classes are equal, and it leads to dense 
packings \cite{Voivret2007}. The total 
number $N$ of particles is 1500. An example of a sample used for simulations 
with a representation of particle pressures is displayed in Fig. \ref{fig:grains}. 

\begin{figure}[tbh]
\centering
\includegraphics[width=0.3\columnwidth]{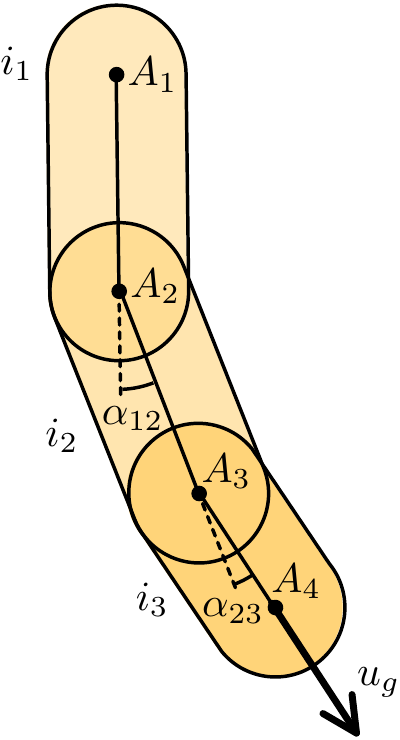}
\caption{(color online) Root particle geometry and connection between consecutive root elements.}
\label{fig:root_elements}
\end{figure}

\begin{figure}[tbh]
\centering
\includegraphics[width=0.6\columnwidth]{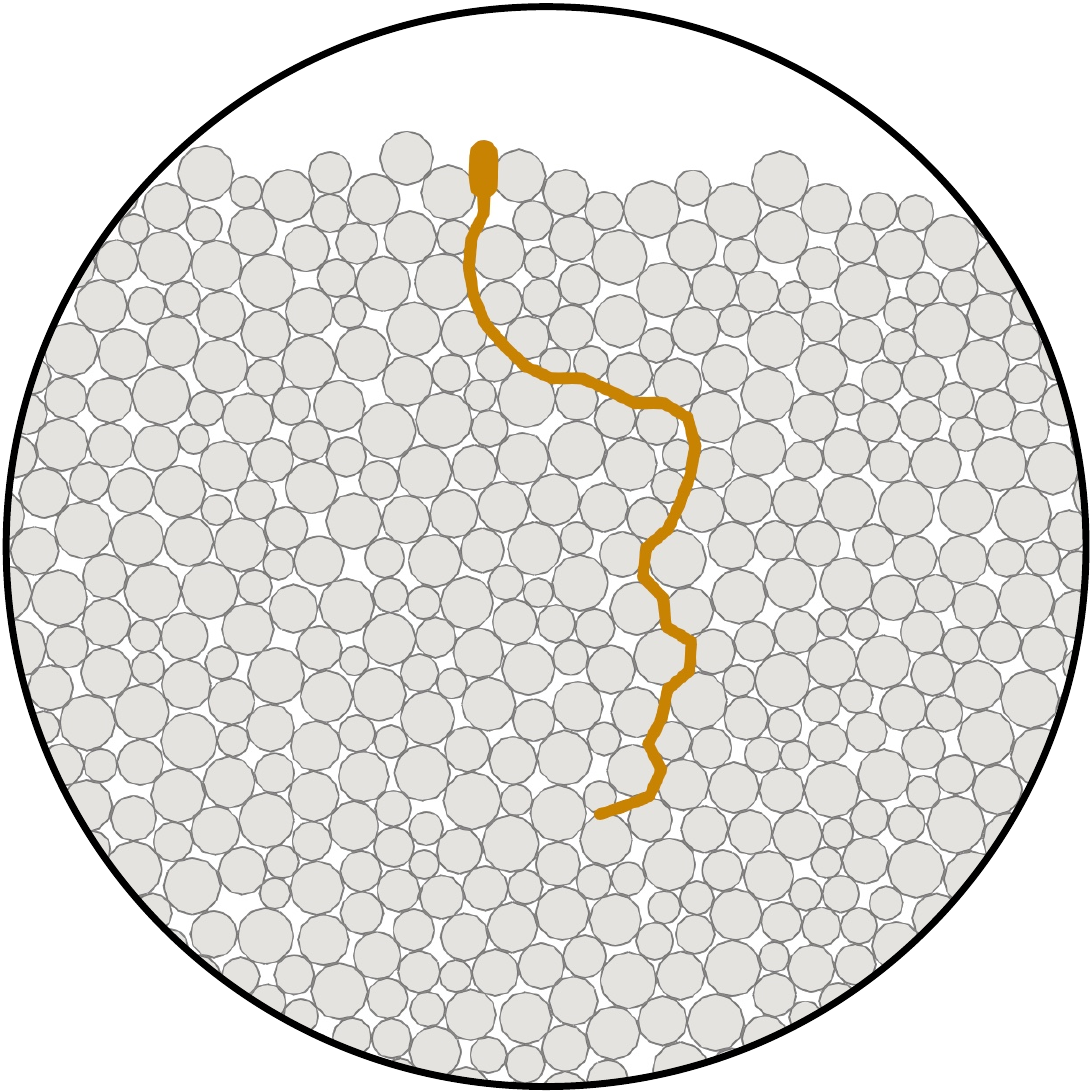}
\caption{(color online) A root grown from a seed placed at the top of a granular sample. }
\label{fig:root_example}
\end{figure}

The root is modeled as a self-elongating spheroline that can change its direction under the action 
of the forces exerted by soil particles. A spheroline is the locus of all points at a given distance 
from a segment, and shrinks to a sphere when the length of the segment tends to zero. 
It can also be described as a rounded-cap rectangle. 
We begin with a seed defined by an immobile disk of radius $R_r$  
placed on top of the sample. The initial direction of growth $\alpha_1$ is assumed to be vertical. 
We make the root grow from this seed by replacing the disk by a spheroline of radius $R_r$ defined  
by a segment $A_1A$, where $A$ is a new point added vertically below $A_1$; see Fig. \ref{fig:root_elements}.   
The root keeps growing by incremental displacement of the point $A$ 
at a constant rate $u_g$. As long as the growth distance $A_1A$ is below a 
maximum length $\ell_g$, the spheroline keeps its direction $\alpha_1$. 
When $A_1A = \ell_g$, the end point $A \equiv A_2$ of the spheroline becomes 
the seed for the growth of a new spheroline in a new direction $\alpha_2$ determined 
by the forces acting on it; see Fig. \ref{fig:root_elements}. 
The same process is repeated with the new spheroline until the length of the new 
root element reaches $\ell_g$. By iterating this process, we obtain 
a sequence of points  $A_1A_2A_3\ldots$ along successive directions $\alpha_1\alpha_2\alpha_3\ldots$ which 
describes the geometry of a root of thickness $d_{root}=2R_r$ grown from the seed $A_1$. Fig. \ref{fig:root_example} 
shows an example of a root  obtained by this procedure. The ratio $\ell_g/R_{min}$ 
controls the precision with which the shape of the root is defined. Below, we refer to the spherolines 
as `root elements'.     

The flexibility of the root depends on the bending (or torsional) resistance 
between consecutive root elements. The endpoint of each root element $i$ is connected 
to the initial point of the next root element $i+1$ by two linear springs. 
The elongational spring connecting the two points exerts a force ${\vec f}^{i(i+1)}$ 
proportional to their separation distance with a longitudinal stiffness $K_e$. 
In the absence of all other forces, ${\vec f}^{i(i+1)}$ vanishes when the two points 
coincide. The stiffness $K_e$ defines the axial stiffness of the root. 
The two elements are also connected by a linear torsional spring  
with a torque $M^{i(i+1)}$ proportional to the difference $\alpha^i - \alpha^{i+1}$ between their directions:
\begin{equation}
M^{i(i+1)} =  K_b \alpha^{i(i+1)} = K_b (\alpha^i - \alpha^{i+1}),
\label{eqn:MKb}
\end{equation}
where $K_b$ is the bending stiffness of the root. 

The motion of the center of mass $\vec r^i$ of each  
root element or soil particle $i$ along the $x$ and $y$ directions is governed 
by the equations of motion:
\begin{equation} 
m^i \, \ddot{\vec r}^{\, i} = \sum_{j\in\Omega^i} \vec f^{\, ij},   
\end{equation}  
where $m^i$ is the mass, and the summation runs over the set $\Omega^i$ of all 
neighboring particles or root elements $j$ exerting a force $\vec f^{ij}$ on particle 
or root element $i$.    
The angular motion of each root element $i$ is governed by the moment equation 
\begin{equation}
I^i \ddot{\alpha}^{i} = M^{i(i-1)}+M^{i(i+1)}+ \sum_{j \in \Omega^i} \vec c^{\, ij} \times {\vec f}^{\, ij}, 
\label{eq:MI}
\end{equation}
where $I^i$ is the moment of inertia of the root element with respect to 
its center of mass, the torques $M^{i(i-1)}$ and $M^{i(i+1)}$ are given by equation (\ref{eqn:MKb}), 
and  ${\vec c}^{ij}$ is the {\em contact vector} joining the center of mass of element $i$ to its contact point 
with particle or element $j$. The angular motions of the particles are governed by the same equation 
without the torque terms.    

The above model of a `discrete' growing root implies that a root 
emanating from a seed will grow along a straight vertical line in the absence of 
soil particles. In the presence of soil particles, the root will frequently meet the particles 
on which it can exert a force, leading either to the displacement of the particles or 
elastic deformation of the root itself. Because of elasticity, however, the root resumes its 
reference shape as a straight line if the soil particles are removed. This is not what one observes 
in real roots, which, unless damaged, keep practically their shape when removed from a soil. 
This is due to the remodeling of the root during its growth as a result of the accretion of new wood material 
at the surface and/or a change in material internal properties \cite{Guillon2012}, and leads to  
two different effects: 1) stiffening of the root and 2) transformation of elastic reorientations of the root 
into permanent deflections. To account for these evolutions of the root, we assume that  
the bending stiffness $K_b$ between the last 
growing root element, which represents the meristem, and the element to which it is attached 
is much lower than the stiffness $K'_b$  of preceding elements. In this way, the tip of the root 
can more easily probe the pore space between soil particles than the  
upstream parts of the root. 

Moreover, we allow the reference equilibrium angle $\alpha_0^{i(i+1)}$ between 
consecutive root elements to evolve. The initial reference angle is zero according to equation (\ref{eqn:MKb}), 
corresponding to a zero torque when the two root elements are co-linear.   
As the root grows, at the same time as a new root element is created, 
we set all reference angles between all root elements, in exception to that between the new element and 
its preceding element, to their current values. As a result, the elastic torque acting between all those elements 
is reset to zero, and the torque in the following steps of computation is given by
\begin{equation}
M^{ij} =  K'_b (\alpha^{ij} - \alpha_0^{ij}),
\label{eqn:MKb_alpha}
\end{equation}   
It should be noted that, although the re-orientations of the meristem are much 
easier than other parts of the root, this evolution of parameters 
does not prevent from later variations of the angles as they still can 
change by elastic deflections depending on the forces exerted by 
the soil particles.       

\begin{figure}[tbh]
\centering
\includegraphics[width=0.5\columnwidth]{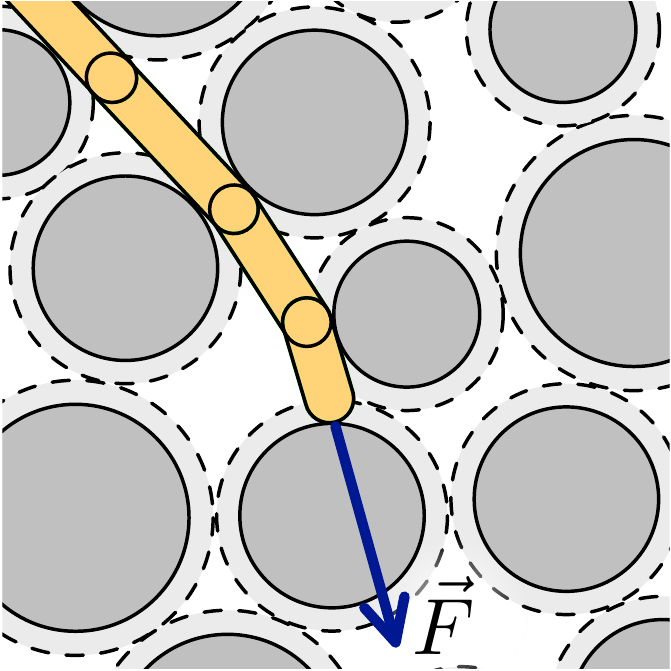}
\caption{(color online) A growing root and the interparticle gaps seen by the root.}
\label{fig:gaps}
\end{figure}

Another aspect that needs to be considered in 2D simulations is 
that, in contrast to 3D packings, all pores in a dense 2D packing are closed, 
implying that the root can not grow without dislodging the particles. In order to relax this 
2D pathology, we use two different radii 
for the particles: one radius $R_p$ for particle-particle contacts and a slightly smaller radius  
$R'_p =R_p - 0.5 d_{gap}$ for the particle-root contacts. Hence, when two particles are in contact, the root `sees' 
a gap of width $d_{gap}$ between the two particles through which it can pass without touching the particles 
if the root diameter $d_{root}$ is smaller than $d_{gap}$, as shown 
in Fig. \ref{fig:gaps}. The ratio $s=d_{root}/d_{gap}$ is a model parameter that plays the same role as 
the ratio of the root diameter to the pore size in 3D and, as we shall see, it will influence the force required for the 
root to dislodge the particles. 

As to the friction coefficients between root elements and particles, we 
consider the plausible assumption that the friction coefficient of soil particles with  
the meristem is smaller than with other root elements because the sloughing of cells from 
the root cap produces a low-friction disposable sleeve \cite{Bengough97}. 
In the simulations reported in this paper, we set the friction coefficient between the root cap and particle to zero. 
The friction coefficient between particles and with other root elements was set to $\mu=0.4$.   

We used the velocity-Verlet stepping scheme to solve the equations of motion 
for all particles and root elements \cite{AllenTildesley1989}. 
The solver was also optimized by using the cell method for the book-keeping of 
close neighbors \cite{meffeja2012}. We set $K_n=K_t=10^{-6}$ N/m and performed a large 
number of simulations with different values of bending stiffness $K_b$, elementary 
growth length $\ell_g$, particle sizes $R_{min}$ and $R_{max}$, gap parameter $s$ and  
packing fraction $\Phi$ of the soil. With each set of parameter values, we performed 
five independent simulations by changing the position of the seed on top of the sample 
in order to evaluate the variability due to granular disorder. 

In all simulations, the growth rate $u_g$ 
was set to a small value in order to avoid inertial effects that are basically absent from 
the very slow growth process of roots. The typical growth time $\tau$ is the time necessary 
for the root tip to grow over a distance of the order of an average soil particle diameter:
\begin{equation}
\tau = \frac{\langle d \rangle}{u_g}
\end{equation} 
This time should be compared with the time necessary for a particle dislodged  
from its equilibrium state to relax back to equilibrium. The typical distance being of the order of one 
particle diameter, the characteristic relaxation time is
\begin{equation}
t_{g} = \left( \frac{\langle d \rangle}{g} \right)^{1/2}, 
\end{equation}
where $g$ is the gravity. Hence, for a quasi-static growth and high time resolution we imposed 
$\delta t \ll t_g \ll \tau$, where $\delta t$ is the simulation time step. 

\section{Root-particle force distributions} 
\label{sec:force}

\begin{figure}
\centering
\includegraphics[width=1\columnwidth]{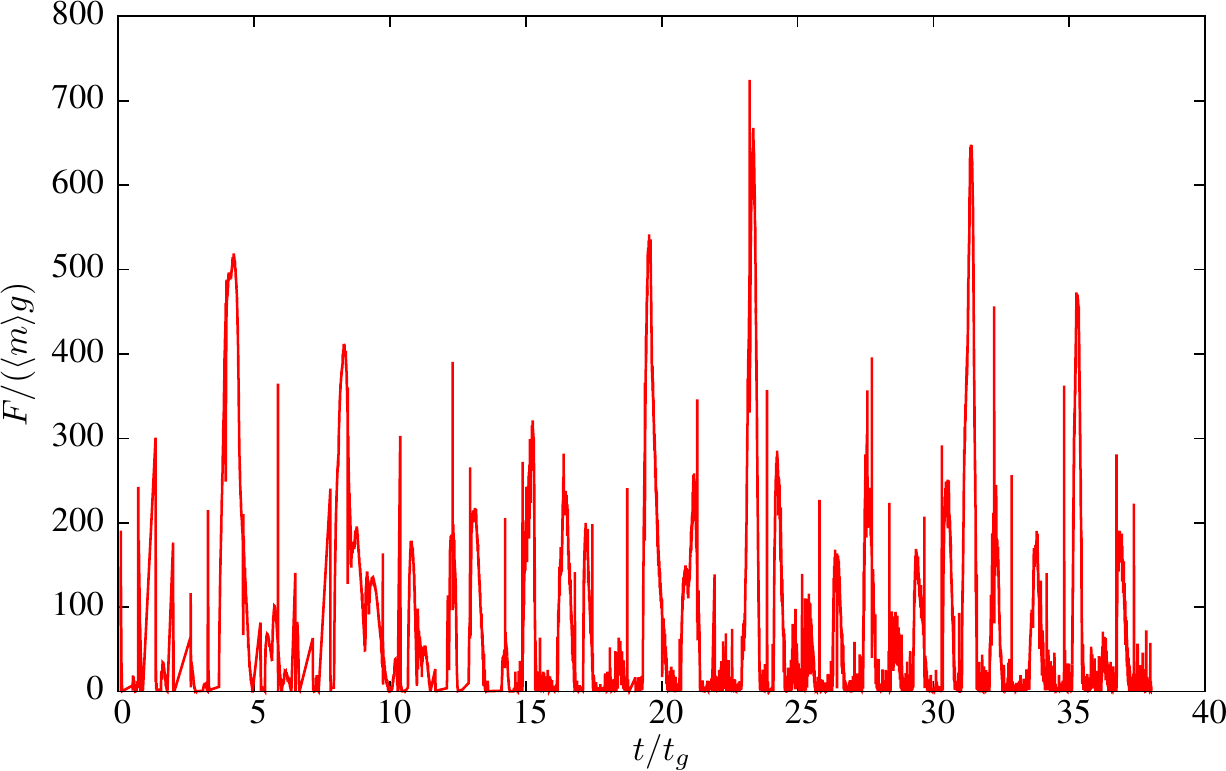}
\caption{(color online) Evolution of the force $F$ experienced by the root cap during its growth 
as a function of normalized time. The force 
is normalized by the mean particle weight.}
\label{fig:fxysignal}
\end{figure}

During the root growth, the positions $x$ and $y$ of the root tip define 
its trajectory. We consider here the total force exerted by the soil particles 
on the root cap, i.e. the force acting on the growing root element. 
Its projection along this root element is the `axial force' $F$ that resists  
root growth; see Fig. \ref{fig:gaps}.    
Fig. \ref{fig:fxysignal} displays the evolution of $F$ with time for a root of bending stiffness 
$K_b = 1$ Nm and $s = 1.54$ in a dense sample of 1500 particles. 
The force is normalized by the mean weight $\langle m \rangle g$ of a single particle whereas 
the time is normalized by the growth time $t_g$. 
We see that the force signal undergoes large variations between zero and values as 
high as 700 times the mean particle weight nearly independently of its depth inside the 
soil. The zero force values correspond to the passage of the root in the pores without contact 
with particles whereas high values occur when the root meets the particles balanced by strong 
force chains. The strong forces in granular materials can be much larger than 
the mean force \cite{Radjai1998}. Although the mean force increases with  
depth, this increase is negligible compared to the large force fluctuations arising from 
root-particle interactions.

\begin{figure}[tbh]
\centering
\includegraphics[width=1\columnwidth]{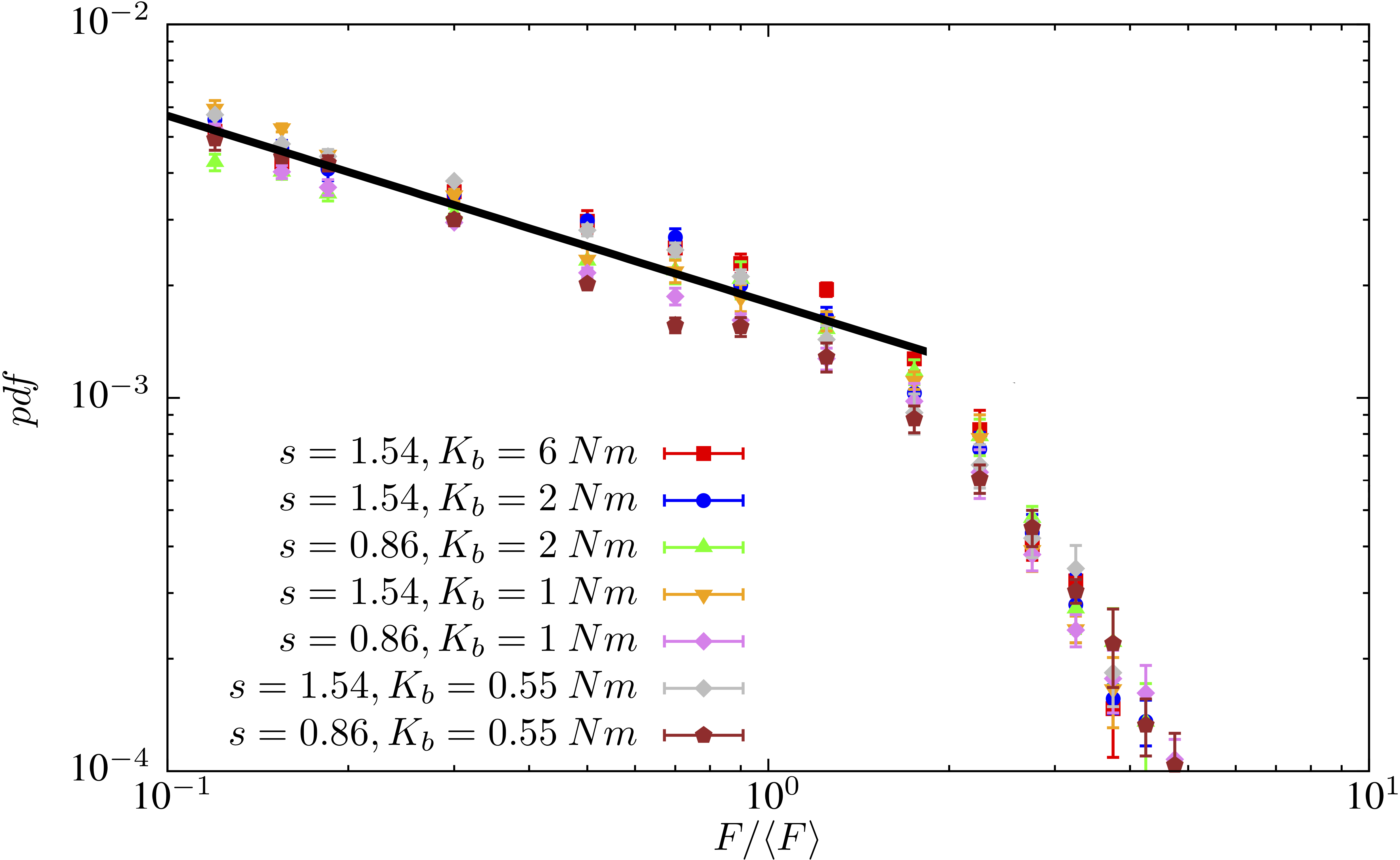}(a) 
\includegraphics[width=0.99\columnwidth]{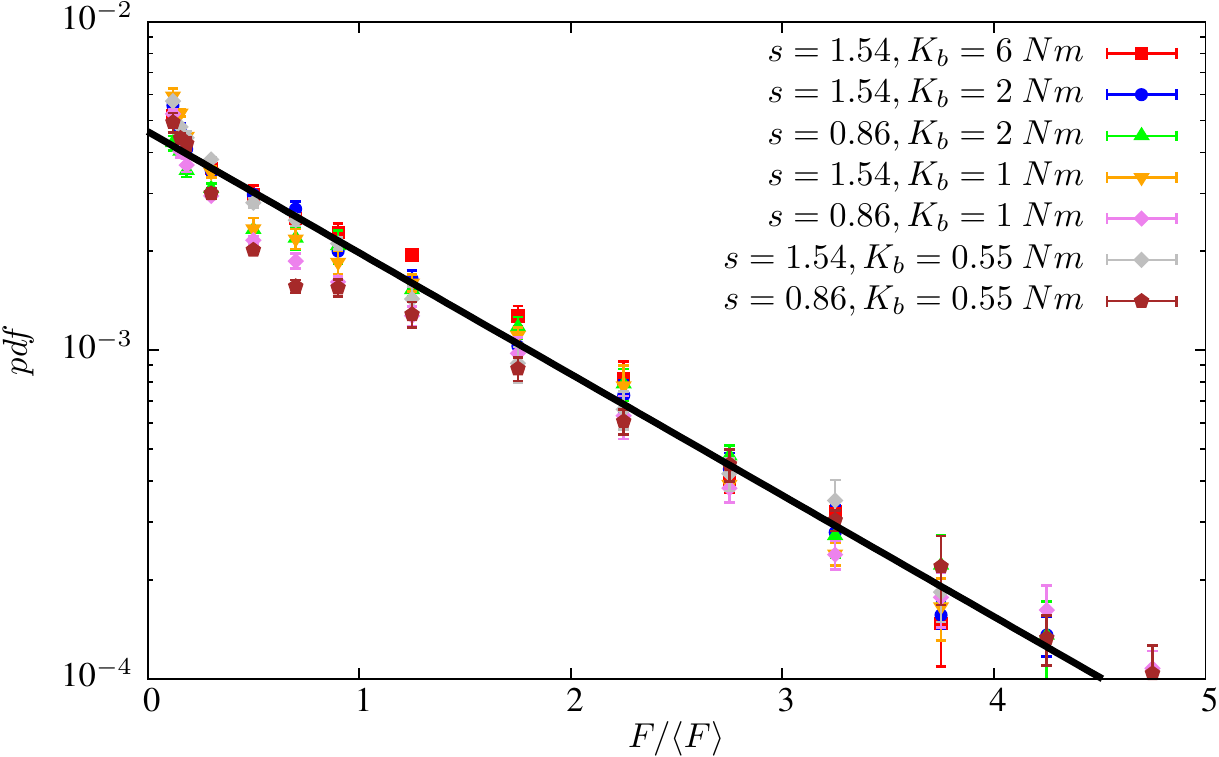}(b) 
\caption{(color online) Probability density function of forces  
experienced by a growing root in a granular packing in log-log and log-linear scales 
for different values of the gap parameter $s$ and bending stiffness $K_b$ of the root.}
\label{fig:hist-force}
\end{figure}

Figure \ref{fig:hist-force} shows the probability density function (pdf) 
of root-particle forces on log-log and log-linear scales 
for two different values of $s$ (above and below 1), 
and for a set of values of $K_b$ ranging from $0.55$ Nm to 6 Nm. 
We see that, despite the broad range of the values of $K_b$, all the data points 
collapse nicely on the same curve when the forces 
are normalized by the mean force $\langle F \rangle$. The small deviations 
reflect the packing disorder and finite system size.       
The distribution of forces below the mean force (weak forces) is a power law $F^{-\alpha}$ 
with exponent $\alpha \simeq 0.5$ over one decade. The distribution of forces above the mean force 
(strong forces) is well fitted by an exponential falloff $e^{-\beta F/\langle F \rangle}$ with 
parameter $\beta \simeq 0.85$. 

This distribution with its transition from a decreasing power-law behavior to an 
exponential falloff around the mean force is reminiscent of the 
pdf of contact forces in granular materials \cite{Liu1995a,Radjai1996,Majmudar2005,Radjai2015}. 
This `bimodal' distribution can be approximated by \cite{Radjai1999a,Richefeu2009,Voivret2009}
\begin{equation}
P(F) = 
\left\{
\begin{array}{lcl}
k \ \left(\frac{F}{\langle F \rangle}\right)^{-\alpha}
\qquad \quad &&N/\langle F \rangle < 1, \\ 
k \ e^{\beta\left(1- F/\langle F \rangle \right)} \qquad 
\quad &&F/\langle F \rangle > 1, 
\end{array}
\right.
\label{eqn3}
\end{equation}         
where  $k$ is the normalization factor given by
\begin{equation}
\frac{1}{k} = \frac{1}{1-\alpha}+\frac{1}{\beta}.
\label{eqn4}
\end{equation}       
Moreover, considering the mean force $\langle F \rangle$ as the 
point of crossover between the two parts of the distribution, we get the following relation between 
the exponents: 
\begin{equation}
\beta^2=(1-\alpha)(2-\alpha).
\label{eqn5}
\end{equation}
This relation is in excellent agreement with the values of the exponents $\alpha$ and $\beta$ 
measured from Fig. \ref{fig:fxysignal}. Although the values of these exponents for 
contact force distributions in granular materials are different 
(lower values of $\alpha$ and larger values of $\beta$ depending on the soil properties 
such as size polydispersity \cite{Voivret2009}), the similarity between their pdf's is a strong indication that 
a growing root probes the weak and strong force networks.     
Their independence from the root bending stiffness 
is consistent with this assumption. 

\section{Effect of the root bending stiffness}
\label{sec:bending}

Although the force pdf is independent of the root and soil characteristics, 
the mean force $\langle F \rangle$ exerted on the root cap during its growth is sensitive to 
the root bending stiffness $K_b$ and soil parameters such as packing fraction $\Phi$, 
mean particle diameter $\langle d \rangle$, and cohesive forces $f_c$ between particles. 
Herein, we consider the effect of $K_b$ as a major root parameter that controls the ability of the root to change its 
direction and explore the pore space. With low bending stiffness the root is more flexible and tends to follow the 
tortuous space of the pores and gaps between particles whereas for large bending stiffness the root 
will be able to dislodge the particles and will therefore follow a path close to a straight line. 

\begin{figure}
\centering
\includegraphics[width=0.9\columnwidth]{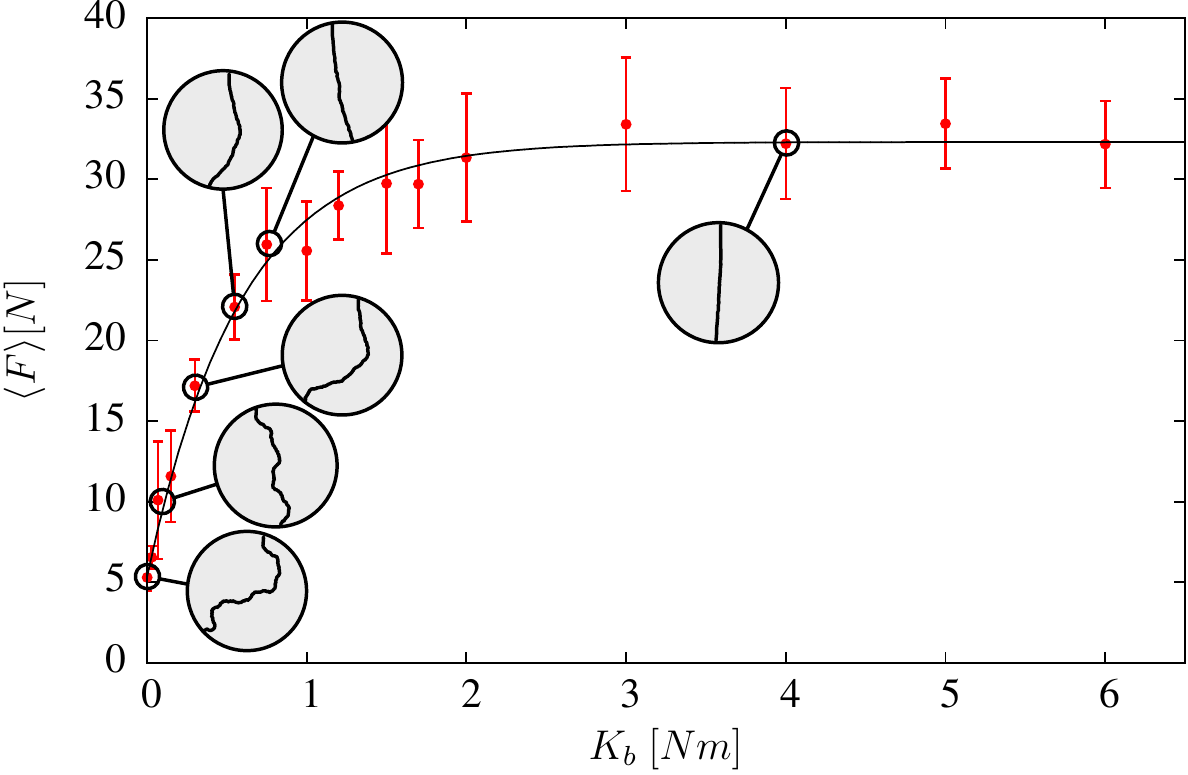}
\caption{(color online) The mean force $\langle F \rangle$ experienced by the root cap as a function of 
bending stiffness $K_b$ for $s=0.86$. The error bars represent standard deviation for 
five independent simulations. The circular insets display a portion of the root for several values of $K_b$.}
\label{fig:Mean-Force}
\end{figure}

Figure \ref{fig:Mean-Force} shows $\langle F \rangle$ as a function of $K_b$ for $s=0.86$. 
We see that the mean force increases from a weak value $F_0$ at very low values of $K_b$ 
and levels off to a constant value $H$ at large values of $K_b$.  
For each value of $K_b$, the data point represents the average force obtained from 5 independent simulations in which only the 
position of the seed is changed on top of the sample. The error bars represent the corresponding 
values of the standard deviation. Within statistical precision, the evolution of  $\langle F \rangle$  with $K_b$ 
is well-fitted by an exponential function  
\begin{equation}
\langle F \rangle = F_0+ (H-F_0)(1-e^{-a' K_b}),
\label{eqn:FKb}
\end{equation}  
where the parameter $a'$ has the dimension $N^{-1}m^{-1}$. 
In the two limits of very high and low values of $K_b$, the force is independent 
of $H$. At high values, $\langle F \rangle = H$ coincides with the resistance of the granular medium to the 
penetration of a rigid bar whereas at low values the force $\langle F \rangle = F_0$ 
results from the collisions of the root with the soil particles and its re-orientations are 
not prohibited by the root bending stiffness. As we shall see below, we find $F_0 \simeq 0.1 H$. 
In Fig. \ref{fig:Mean-Force} we also have displayed a portion of the root at different steps of its 
growth with a shape that evolves gradually from a random walk in the pore space to a straight line 
as $K_b$ increases.

Equation (\ref{eqn:FKb}) is interesting as it describes in a simple and quantitative way the effect of 
root flexibility on the mean force involved in root growth. 
This relation makes it possible to predict the mean force exerted
on the root cap during it growth from the
knowledge of the root stiffness and the soil hardness $H$ 
defined as the resistance of the soil to the penetration of a rigid bar as in penetrometer experiments 
\cite{Bengough91,QuezadaSaussineBreulEtAl2014}. Equation (\ref{eqn:FKb}) 
shows clearly that the mean force experienced by the root cap can actually be 
much lower than $H$ as measured from the ground by a penetrometer 
depending on the root flexibility. It should be noted here that the longitudinal 
and tangential stiffnesses of the root are set to a high value, 
and we checked that their values are not significant for the results presented 
below.   

Equation (\ref{eqn:FKb}) 
with its values of $H$ and $F_0$ was obtained for specific values of soil parameters 
such as particle size distribution, gap size and packing fraction without 
cohesive forces acting between soil particles. 
In order to examine the robustness of equation (\ref{eqn:FKb}) under more general 
conditions and investigate the influence of relevant parameters,  
we performed extensive simulations of root growth with a range of 
the parameter values.  

\section{Influence of soil parameters}
\label{sec:soil}

We first consider here the effect of the packing fraction $\Phi$, which is 
an important soil property that depends on material parameters such as 
friction coefficient $\mu$ between particles but also on the history of 
soil deformations. As we would like to single out the effect of the packing fraction, 
we fix all material parameters but apply a preparation method to obtain different 
values of $\Phi$.         

\begin{figure}[tbh]
\centering
\includegraphics[width=0.9\columnwidth]{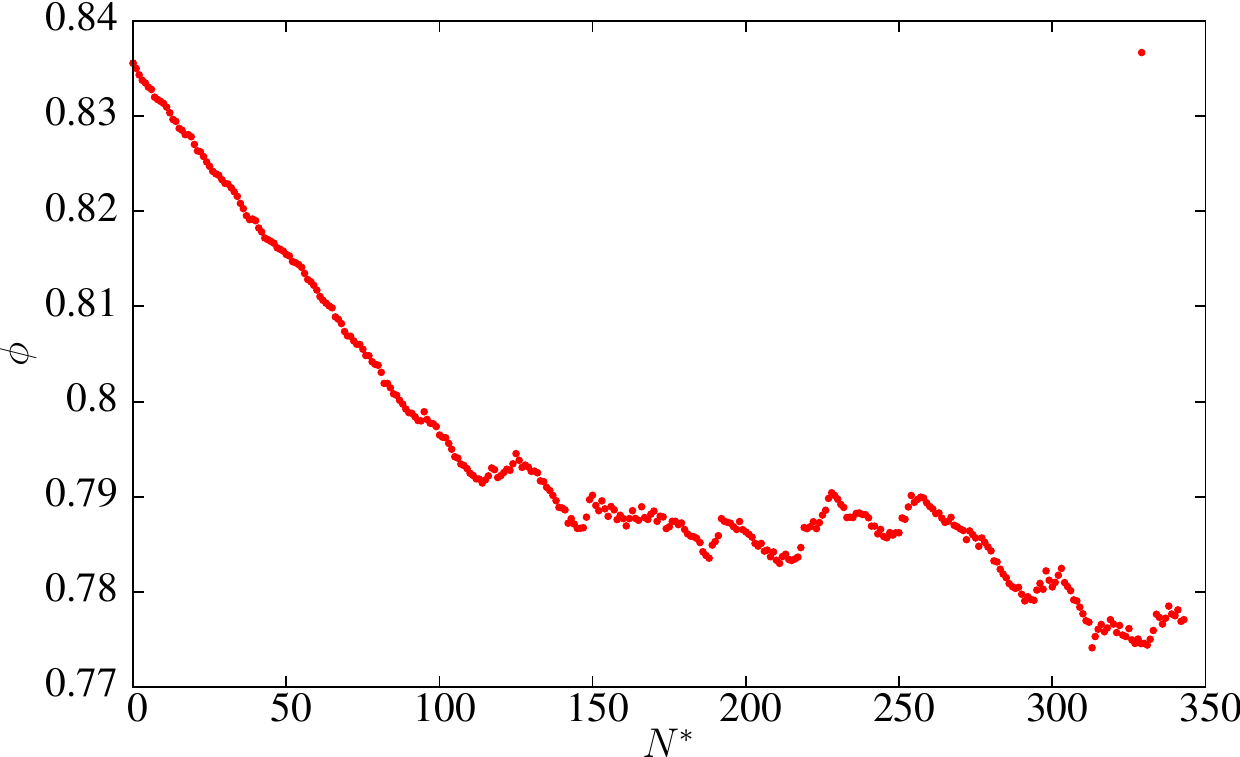}
\caption{(color online) Variation of packing fraction $\Phi$ as a function of the number $N^*$ of particles 
removed.}
\label{fig:compacity}
\end{figure}

To prepare samples with different values of the packing fraction, we first prepare 
a dense packing of 2000 particles by setting $\mu$ to zero and allowing the particles fall into the simulation 
box from their initially random positions in space. The zero value of $\mu$ reduces the 
probability of arching and leads to  a dense packing with $\Phi=0.8355$. 
Then, we set $\mu=0.4$ and remove one particle at a time randomly selected from  
the packing. After each removal, we allow the packing to relax towards equilibrium 
over several time steps.  The equilibrium criterion is chosen according to
to the particle velocities and in order to make the process numerically efficient.
Thus, a new particle is removed when the mean particle velocity in the sample is much lower than  
$\sqrt {g \langle d \rangle} $, which is the falling time of a particle of average diameter $\langle d \rangle$ over 
a distance equal to its diameter. 
Removing a particle creates a local imbalance 
and therefore entails a rearrangement of the particles. But this rearrangement is not necessarily local. 
As a result, $\Phi$ decreases by removal but  increases again 
by relaxation. Since we begin with high packing fraction, the removal-relaxation 
process leads to a gradual decrease of $\Phi$ until a statistical balance is achieved between the
phases of decrease and increase of $Phi$, as shown in Fig. \ref{fig:compacity}. The average packing 
fraction in this steady state is $\simeq 0.78$ after the removal of 350 particles.   

\begin{figure}[tbh]
\centering
\includegraphics[width=1\columnwidth]{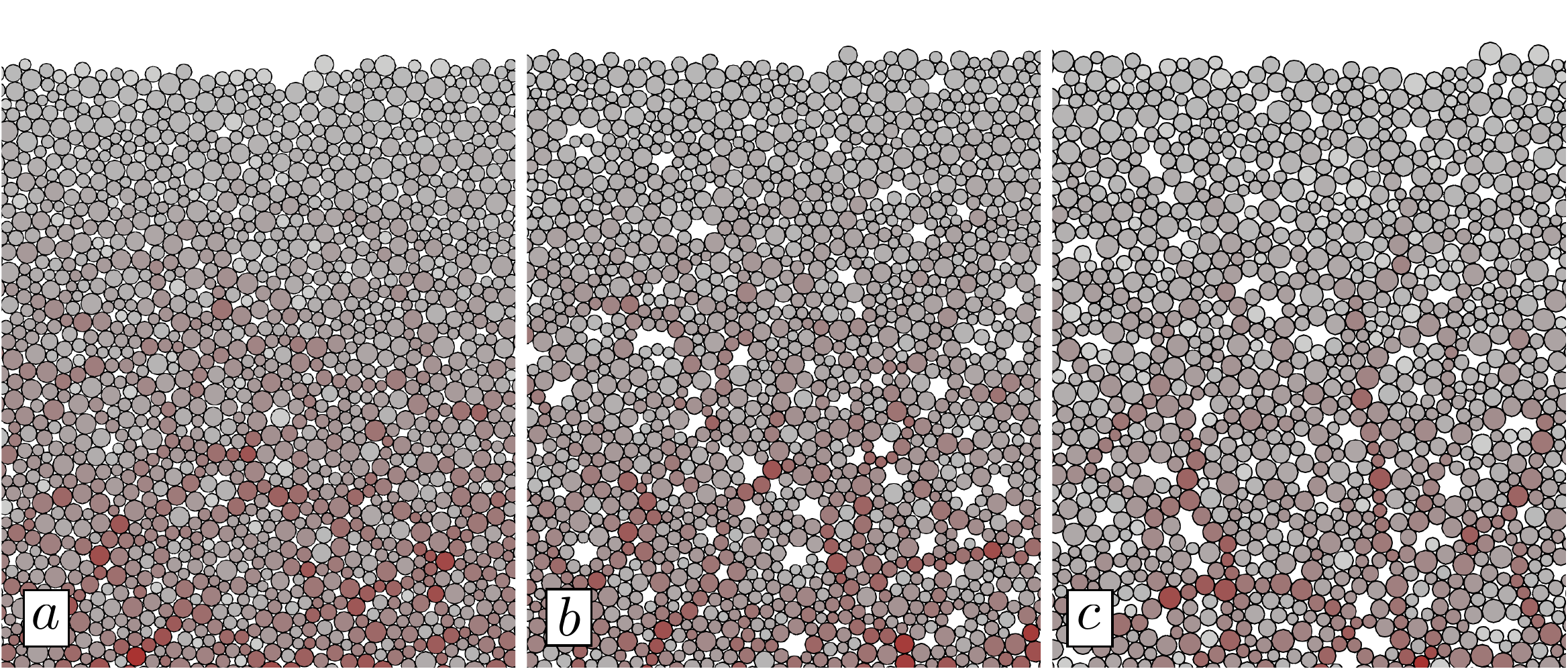}
\caption{Samples of different packing fractions obtained by a removal-relaxation process. 
Line thickness is proportional to normal force between particles.} 
\label{fig:solcomp}
\end{figure}

\begin{figure}[tbh]
\centering
\includegraphics[width=0.9\columnwidth]{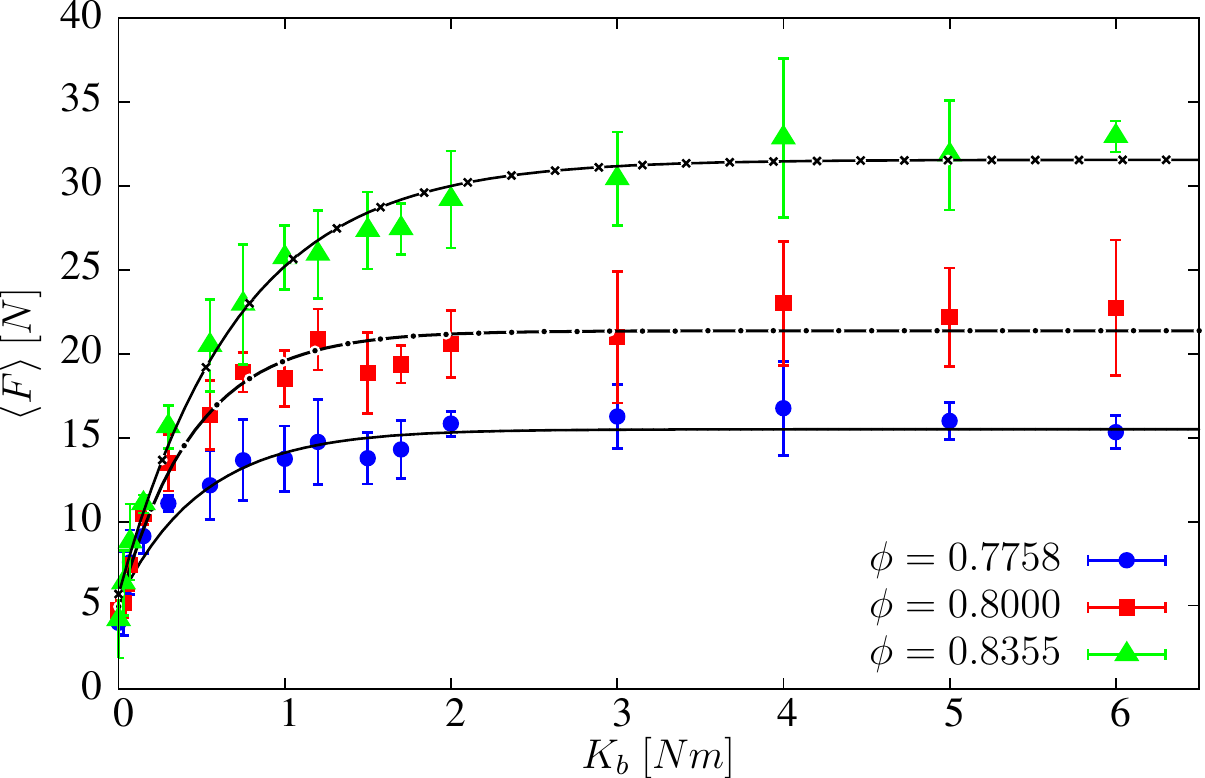}(a)
\includegraphics[width=0.935\columnwidth]{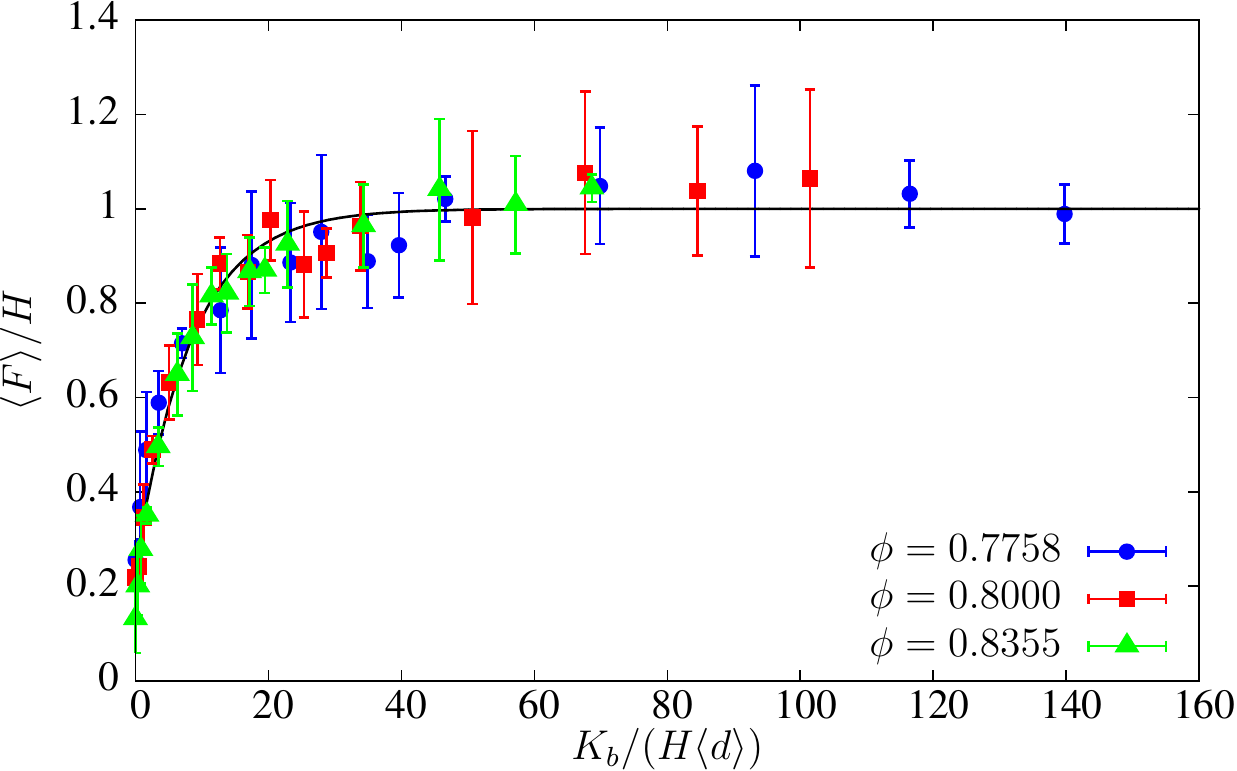}(b)
\caption{(color online) (a) The mean force as a function of the root bending stiffness for 
three samples of different packing fractions; (b) The same data points with the mean force normalized 
by soil hardness $H$ and bending stiffness normalized by $H$ times the mean 
particle size $\langle d \rangle$. The lines are exponential fits according to equation (\ref{eqn:FKb}) 
in (a) and equation (\ref{eqn:F3}) in (b).}
\label{fig:comp}
\end{figure}

We selected three samples of packing fractions 0.77, 0.8 and 0.83, displayed 
in Fig. \ref{fig:solcomp}. Despite locally large variations of porosity, we checked that there is no 
porosity gradient by dividing each sample to eight horizontal layers. Then, we carried out 
root growth simulations for increasing values of $K_b$ in each sample.  
Fig. \ref{fig:comp}(a) shows the mean force as a function of $K_b$ for the three samples.  
The asymptotic force $H$ increases considerably
with $\Phi$. But, within statistical precision, the data are well fit by 
the exponential form of equation (\ref{eqn:FKb}), and thus 
for each value of $K_b$ the mean force increases with packing fraction. 

The exponential fit of equation (\ref{eqn:FKb}) clearly suggests that the mean force 
should be scaled by $H$ but this normalization of the forces does not ensure that they will collapse 
on the same curve as the values of the prefactor $a'$ in the exponent are different 
in the three exponential fits. Hence, we also need to scale $K_b$ by an appropriate 
material parameter. $K_b$ has the dimension of a force multiplied by a length.
$\ell_g$ being a model discretization parameter, the only characteristic length 
of the packing is the average particle size $\langle d \rangle$. 
We also have two characteristic forces: 1) the mean particle weight $ \langle m \rangle g$ and
2) the hardness $H$. As previously discussed, the root-particle forces are considerably 
larger than  $ \langle m \rangle g$ and they result from both the interparticle force chains 
and the root-particle interactions. The only force reflecting both these aspects is $H$. 
This suggests that we should normalize the bending stiffness by $ H \langle d \rangle $. 
As observed in Fig. \ref{fig:comp}(b),  with this scaling all the data points from the three samples 
nicely collapse on a single curve well fit by
\begin{equation}
\frac{\langle F \rangle}{H} = 1 - \left(1 - \frac{F_0}{H} \right) e^{-a K_b/(H \langle d \rangle)}, 
\label{eqn:F3}
\end{equation}        
where $a \simeq 0.15$ and $F_0/H \simeq 0.1$. This scaling means that the effect of the packing 
fraction $\Phi$ is fully captured by the dependence of $H$ on $\Phi$, and from its value  
the mean force exerted on the root cap  can readily be deduced by means of equation (\ref{eqn:F3}). 

In the data represented in Fig. \ref{fig:comp}, $H$ takes three different values 
but $\langle d \rangle$ is the same in all samples. In order to 
further validate the scaling proposed in (\ref{eqn:F3}), we performed root growth 
simulations for several values of $R_{min}$ and $R_{max}$.      
In the simulations presented in the previous section, 
we had $\langle d \rangle=3.1$ mm for 1500 grains with radii varying 
between $ R_{min}=1 $ mm and $ R_{max}=2 $ mm. 
We prepared two more samples of mean particle diameters $ 1.6 $ mm and $ 4.2 $ mm.  
In the first case we reduced $R_{min}$ and increased     
the number of particles to 2000 to keep 
the same volume as the reference sample. In the second case, we increased 
$R_{max}$ and reduced the number of particles to 1000. The ratio $\lambda = R_{max}/R_{min}$ 
is  2 in the reference sample and, respectively, 2.5 and 5 in the two new samples. 

\begin{figure}[tbh]
\centering
\includegraphics[width=0.9\columnwidth]{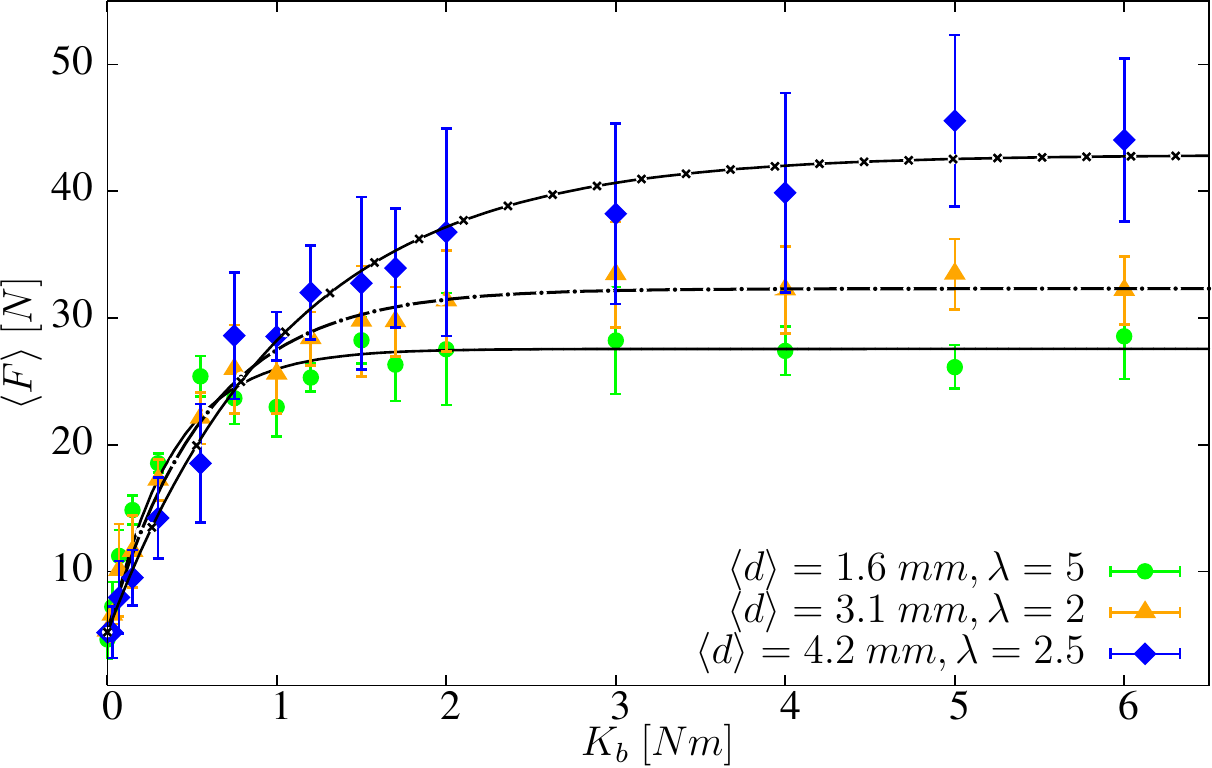}(a)
\includegraphics[width=0.935\columnwidth]{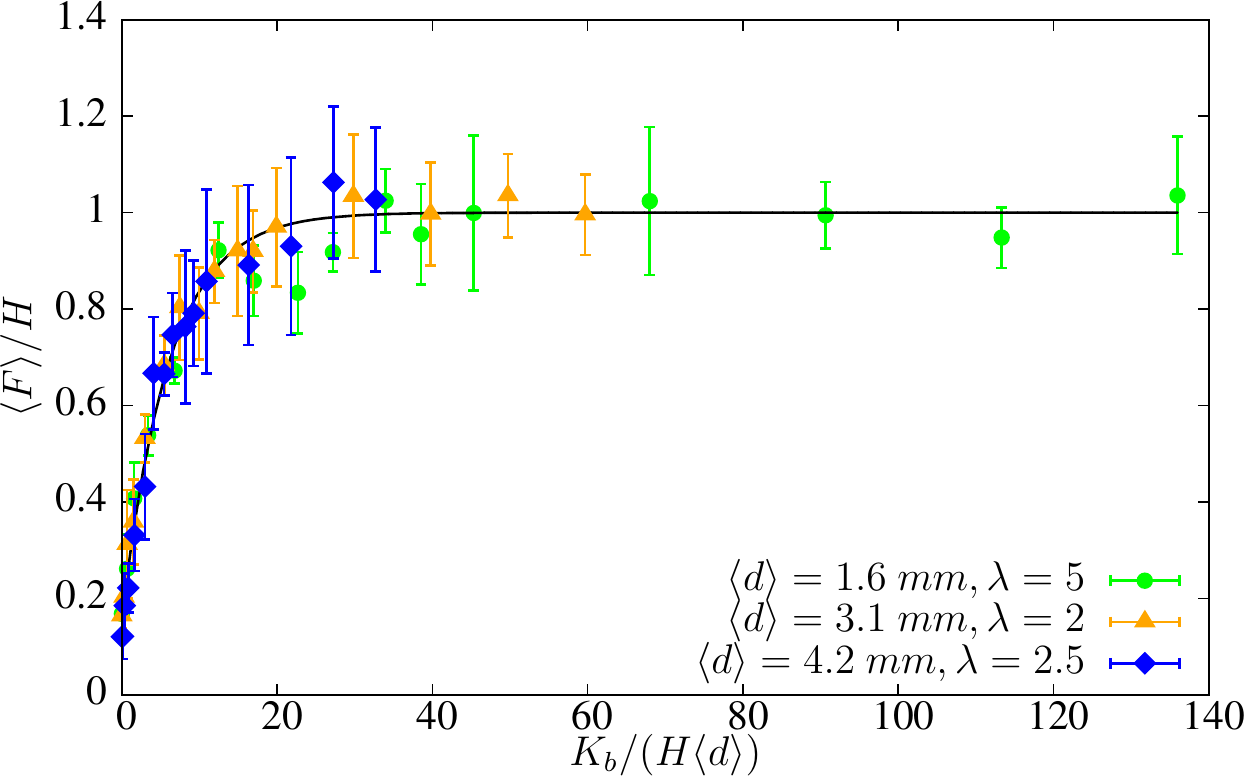}(b)
\caption{(a) The mean force as a function of the root bending stiffness for 
three samples with different values $\langle d \rangle$ and size ratio $\lambda$; 
(b) The same data points with the mean force normalized 
by soil hardness $H$ and bending stiffness normalized by $H\langle d \rangle$. 
The lines are exponential fits according to equation (\ref{eqn:FKb}) 
in (a) and equation (\ref{eqn:F3}) in (b).}
\label{fig:granulo}
\end{figure}

Figure \ref{fig:granulo}(a) shows the mean root-particle force $\langle F \rangle$ as a 
function of $K_b$ for the three samples. Here again, we see that the functional form 
(\ref{eqn:FKb}) provides a nice fit to the data with hardness $H$ increasing considerably 
with $\langle d \rangle$.  This is not surprising
as for high bending stiffness the root dislodges nearly all particles on its path and thus 
the force $H$ directly reflects the contact forces, which for the stress induced by 
particle weights is proportional to $ \langle d \rangle $ in 2D.
The increase of $\langle F \rangle$ is also related to the value of $\lambda$, which 
describes here the polydispersity of the samples. However, $H$ is not an increasing function of 
$\lambda$. This dependence needs to be determined by more simulations. 
Fig. \ref{fig:granulo}(b) displays the same data with normalized values of 
$\langle F \rangle$ and $K_b$. All the data collapse well and can again be  
fitted by the form (\ref{eqn:F3}) with the same value $a \simeq 0.15$ as before, 
and shows once more that the scaling proposed here is robust with respect 
to the soil and root parameters.

We also checked by systematic simulations the influence of the gap parameter 
$s$ and the growth length $\ell_g$. The latter has little effect on the mean force 
$\langle F \rangle$ as far as its value is below $\langle d \rangle$. 
We find that the same scaling applies and the effect of $s$ is captured   
by the increase of $H$ with $s$. As for 
other parameters, more simulations are needed to determine the functional 
dependence of $H$ with respect to $s$.      

\begin{figure}[tbh]
\centering
\includegraphics[width=0.45\textwidth]{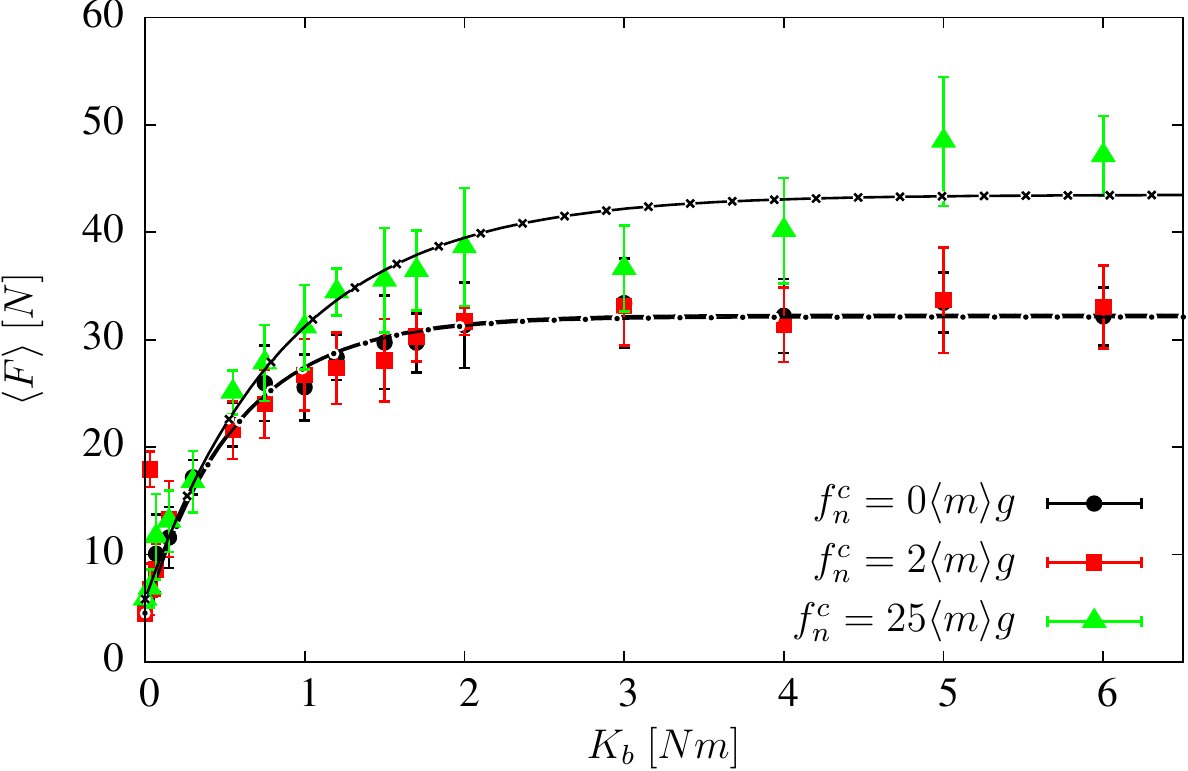}(a)
\includegraphics[width=0.45\textwidth]{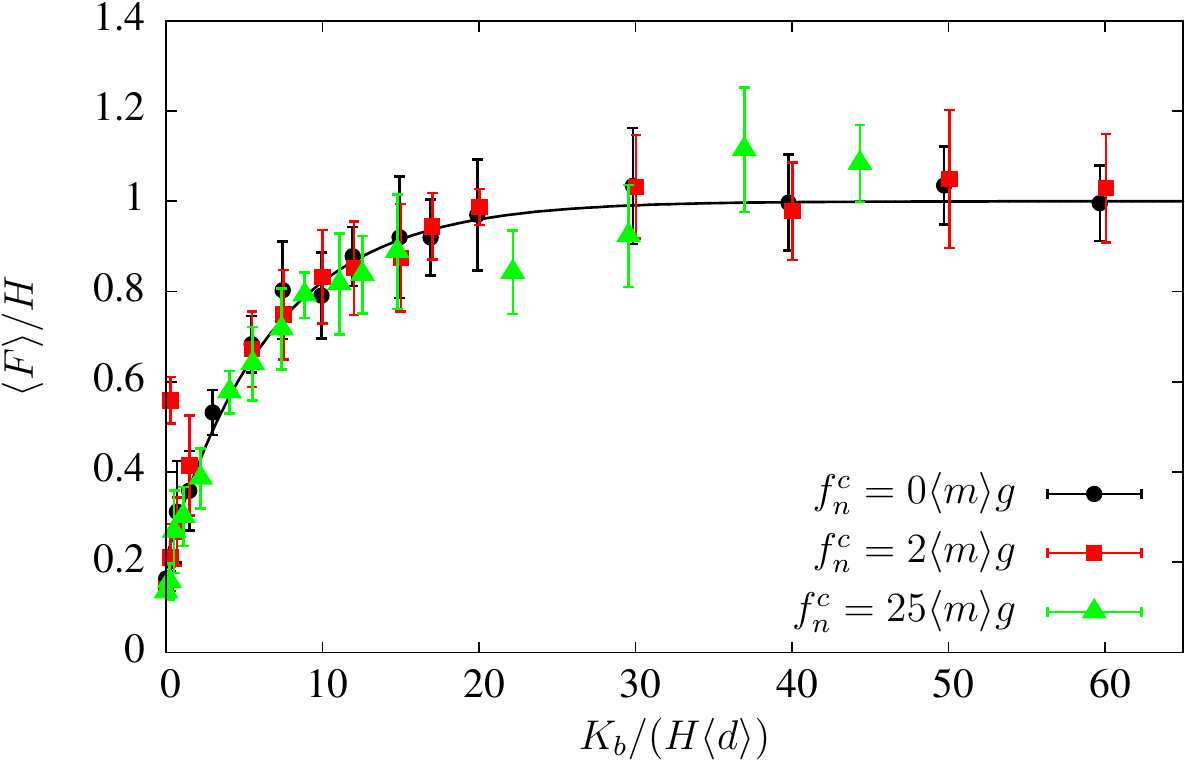}(b)
\caption{(color online) (a) The mean force as a function of the bending stiffness 
for three samples with three values of the adhesion force:  0, 2$\langle m \rangle g$ and 25$\langle m \rangle g$. 
The solid lines are exponential fits; (b) The same data with normalized coordinates. The solid line is 
the functional form of equation (\ref{eqn:F3}).}
\label{fig:cohesion}
\end{figure}

Finally, we investigated the influence of cohesion on the 
root-particle force by assuming that the adhesion force $f_n^c$ in 
equation (\ref{eq:Grains_interaction}) is constant and localized 
at the contact point between the particles whereas the root-particle contacts 
are cohesionless. In the presence of cohesion, the root will have to overcome 
an additional tensile force $-f_n^c$ between the soil particles to advance. 
For the simulations we used the sample  of 1500 grains with particle diameter distribution
between 2 and 4 mm and a mean diameter $ \langle d \rangle $ = 3.1 mm with 
a root diameter larger than the gap between the particles ($ s = 1.54$). 
We used three values of $f_n^c$: 0, 2$\langle m \rangle g$ and 25$\langle m \rangle g$. 
Fig. \ref{fig:cohesion}(a) shows $\langle F \rangle$ as a function of $K_b$ 
for the three samples. The data follow an increasing exponential function as 
in cohesionless samples. The hardness $H$ has nearly the same value for 
$f_n^c =0$ and $f_n^c = 2\langle m \rangle g$ but increases for 
$f_n^c = 25\langle m \rangle g$ as expected. The same data in normalized coordinates 
shown in Fig. \ref{fig:cohesion}(b) collapse on a curve well fit by equation (\ref{eqn:F3}).

As the cohesion due to capillary and clayey forces is a major 
ingredient of soils, this result considerably extends the relevance of 
equation \ref{eqn:F3} with its scaling parameters 
for the analysis of root-particle forces. In practice, the 
effects of cohesion and packing fraction can not be easily 
separated. In the simulations we considered the same sample 
and changed the value of adhesion force. But when a cohesive granular material 
is prepared by pouring the particles into a box, the packing fraction decreases 
as the cohesion increases, so that the combined effect of these parameters may 
cancel out. Such effects need to be investigated  in order to determine how the hardness parameter $H$ 
depends on various soil parameters.

\section{Conclusions}
\label{sec:conclu}

In this paper we introduced a simple discrete numerical model of root growth 
inside a granular soil. This model accounts for the root flexibility and 
its stiffening during is growth. At constant growth rate, the root development 
is guided by the disordered texture of the granular medium and the reaction forces 
exerted by the particles on the meristem, modeled here as the growing 
segment of the root. We found that these forces are well above the mean 
particle weight almost independently of the depth, and their statistical distribution 
reflects that of interparticle forces with an exponentially decaying 
number of strong forces. This feature is independent of the soil characteristics and root 
flexibility once the forces are normalized by the mean force. 

By means of extensive 
simulations, we also showed that the mean root-particle force $F$  
is an increasing exponential function of the root bending stiffness $K_b$ with an asymptotic value $H$ 
(at large values of the latter) that is a function of the soil characteristics such as packing fraction $\Phi$, 
mean particle size $\langle d \rangle$ and cohesion $f_n^c$. 
With a low bending stiffness, the root can change its direction in response to 
a weak contact force exerted by a particle whereas with high bending stiffness the re-orientation of the 
root requires larger contact forces that may dislodge the particle before the re-orientation can occur. 
Hence, the largest force $H$ corresponds to a very stiff root such that it advances along a nearly straight line 
by dislodging all particles on its pathway.                    
More interestingly, our simulations suggest that the functional dependence of 
$F/H$ on $K_b/(\langle d \rangle H)$ is almost universal, the effect of soil characteristics being  
fully captured by the value of $H$. A practical implication of this scaling is that the mean force experienced by 
the root cap during its growth can be estimated from the values of $H$ (measurable by means of penetrometric tests), 
mean particle size $\langle d \rangle$ and root bending stiffness $K_b$.  

The effect of soil parameters on the value of $H$ can be further quantified by 
root growth simulations using a fixed high value of $K_b$ but many more 
soil samples with varying packing fraction, size polydispersity and cohesion should 
be prepared. We analyzed the shape of the developed root from the 
same data, showing strong correlation with the mean force. The results will be published 
elsewhere. Finally, the extensive parametric investigations presented in this paper 
were made possible by the 2D geometry of the system, which requires much less computation 
time and memory than in 3D. But we presently work on an equivalent 3D model 
for comparison with 2D.

From the biological point of view, providing an analytical formulation of the 
reaction forces experienced by a root growing in a heterogenous medium is a 
significant progress as this information is almost impossible to measure in the field. 
Linking this information to a model of root shape, 
which is easy to characterize by image analyses, will also give an indirect way 
to estimate the mechanical history of a growing root. 
This work will finally lay the ground to address the issues of mechanosensing 
and thigmomorphogenesis \cite{Braam2005} in ``the hidden part" of plants.


\acknowledgments{This work was carried out within the framework of the RoSoM project. 
This project is supported by  Agropolis Fondation under the reference ID 1202-073 through 
the Investissements d'avenir program (Labex Agro: ANR-10-LABX-001-01).}


\bibliographystyle{apsrev}


\end{document}